\documentclass[11pt]{article}
\usepackage[margin=1in]{geometry}
\usepackage{amsmath,amssymb,graphicx,booktabs,xcolor,hyperref}
\usepackage[numbers,sort&compress]{natbib}
\hypersetup{colorlinks=true,linkcolor=blue!50!black,citecolor=blue!50!black,urlcolor=blue!50!black}

\title{What crossing symmetry forbids just above $c=1$}
\author{Yu Nakayama\\[4pt]
{\small\itshape Yukawa Institute for Theoretical Physics, Kyoto University}\\[2pt]
{\small\texttt{yu.nakayama@yukawa.kyoto-u.ac.jp}}}
\date{\today}

\begin{document}
\begin{flushright}
{\small YITP-26-142}
\end{flushright}
{\let\newpage\relax\maketitle}

\begin{abstract}
What does a unitary two-dimensional conformal field theory look like just above $c=1$? Using the numerical conformal bootstrap, we study the four-point
function of a scalar Virasoro primary $\phi$ such that no primary lighter than $\phi$ appears in its operator product expansion
with itself. Slightly above $c=1$, crossing symmetry excludes the interval $0.17\lesssim\Delta_\phi\lesssim1$,
although the compact boson realizes every such $\Delta_\phi$ at $c=1$. Combining this exclusion with results from the modular bootstrap, we show that the lightest primary of a compact unitary conformal field theory at
$c=1.0005$ or $1.001$ has dimension $\Delta_{\rm gap}<0.17$. Further from $c=1$, the largest excluded central charge at fixed $\Delta_\phi$ has two sharp minima near
$\Delta_\phi=1/2$ and $4/7$. At these values it lies within $2\times10^{-7}$ of $c=8/7$, where
the $\mathbb{Z}_5$ parafermion and its charge-conjugation orbifold have scalars of exactly these dimensions.
\end{abstract}
\newpage

\section{Introduction}

``Everything not forbidden is compulsory.'' This popular form of Gell-Mann's
principle\footnote{Gell-Mann stated the principle for strong interactions and compared it with a perfect totalitarian
state, where ``Anything that is not compulsory is forbidden''~\cite{Gell-Mann:1956iqa}. The contrapositive of this sentence, the form quoted above, appears in T.~H.~White's \emph{The Once and Future King}~\cite{Kragh:2019tp}.} came to the author's mind
when Simeon Hellerman asked whether there is a compact unitary conformal field theory (CFT) with $1<c<1.001$. The author answered yes, Hellerman
no. The author's reason was that nothing in the representation theory of the Virasoro algebra rules out such a theory,
which should therefore exist.

Below $c=1$, in contrast, representation theory rules out all but a discrete set of theories. Friedan, Qiu and
Shenker showed that unitarity allows only the central charges $c=1-6/m(m+1)$, each with finitely many conformal
weights~\cite{Friedan:1983xq}, so that the unitary theories below $c=1$ are built from the minimal
models~\cite{Belavin:1984vu,Cappelli:1987xt}. Above $c=1$, every conformal weight $h\ge0$ gives a unitary
representation.

Yet no compact unitary CFT with $1<c<8/7$ is known. The known theories closest to $c=1$ from above belong to two
series whose central charge increases through $c=1$, the $\mathbb{Z}_k$ parafermions~\cite{Fateev:1985mm,Gepner:1986hr}
with $c=2(k-1)/(k+2)$ and the $\mathcal{N}=1$ superconformal minimal
models~\cite{Friedan:1984rv,Bershadsky:1985dq,Eichenherr:1985cx,Goddard:1986ee} with $c=3/2-12/m(m+2)$. Both series
reach $c=1$ at $k=m=4$, where the theories are orbifolds of the compact boson~\cite{Ginsparg:1987eb}. The next theories
in the two series are the $\mathbb{Z}_5$ parafermion at $c=8/7$ and the $\mathcal{N}=1$ minimal model with $m=5$ at
$c=81/70$. Together with the charge-conjugation orbifold of the $\mathbb{Z}_5$ parafermion, they are the first known
theories above $c=1$. A compact unitary CFT with $1<c<8/7$ could be neither free nor a tensor product of known
theories, since no free theory and no tensor product of unitary theories has $1<c<6/5$. Because
representation theory allows such a theory, whether it exists is a question about the remaining consistency conditions,
modular invariance and crossing symmetry.

Hellerman himself initiated the modular bootstrap~\cite{Hellerman:2009bu}, in which one bounds the dimension
$\Delta_{\rm gap}$ of the lightest primary as a function of $c$ by modular invariance of the torus partition function.
After later improvements~\cite{Hellerman:2010qd,Friedan:2013cba,Qualls:2013eha,Collier:2016cls}, the bound at $c=1$
is $1/2$, the gap of the compact boson at the self-dual radius~\cite{Collier:2016cls}. With an upper bound on the gap alone, however, one cannot show that no theory exists at a given $c$, because
any theory whose lightest primary lies below the bound is consistent with it. Benjamin, Fitzpatrick, Li and
Thaler instead searched directly for modular-invariant partition functions with $1<c<8/7$~\cite{Benjamin:2026lbj}.
They found candidate spectra with light scalars at $c=1.05$--$1.13$, but none with $1<c<1.06$ and
$0.3<\Delta_{\rm gap}<0.5$. In preliminary work with Langhoff, the same authors also impose integer degeneracies~\cite{Fitzpatrick:2023lvh} and find
that the bound on the gap drops to about $0.2$ near $c=1.01$~\cite{Thaler:2026talk}. Whether their
candidates are actual CFTs, however, cannot be decided from the partition function alone.

Four-point functions carry more information than the partition function, since their crossing equation involves the
coefficients of the operator product expansion (OPE) as well as the spectrum. Several groups have studied the crossing equation with Virasoro or super-Virasoro blocks
numerically~\cite{Chang:2015qfa,Lin:2015wcg,Lin:2016gcl}, and fusion rules have been imposed on the crossing equation
with global conformal blocks~\cite{Nakayama:2025mrm}. At $c>1$, Besken bounded the dimension of the first non-trivial primary in the OPE of
two identical scalars~\cite{Besken:2021eps}. Collier, Kravchuk, Lin and Yin found numerical evidence that a
four-point function of identical scalars whose OPE contains only scalar Virasoro primaries is that of Liouville
theory~\cite{Collier:2017shs}, which is not compact. Poland has recently bounded the average exchanged dimension in four-point functions of identical scalars at
$c=8/7$, $81/70$ and close to $c=1$, imposing only unitarity on the exchanged operators~\cite{Poland:2026zyr}. Several
scalars of the known theories at these central charges lie on or near the boundary of his allowed region. These results
constrain four-point functions at $c>1$, but none of them bounds the dimension of the lightest primary of a compact
unitary CFT just above $c=1$.

In this paper we apply crossing symmetry to the four-point function of a scalar Virasoro primary $\phi$, assuming only that no
primary lighter than $\phi$ appears in the $\phi\times\phi$ OPE. Every compact CFT satisfies this assumption for its
lightest primary, which just above $c=1$ is a scalar (Section~\ref{sec:window}). For each central charge $c$ and each
dimension $\Delta_\phi$, we ask whether the crossing equation of this four-point function can be satisfied with
non-negative squared OPE coefficients. Where it cannot, no unitary CFT has such a primary $\phi$, and we call the set
of these points $(c,\Delta_\phi)$ the excluded region. Using the numerical conformal
bootstrap~\cite{Rattazzi:2008pe,Poland:2018epd}, we keep only the derivatives of the crossing equation up
to total order $\Lambda$, so that the excluded region we find can only grow as $\Lambda$ increases. We pay particular
attention to sharp features of the region that remains allowed, because in higher dimensions known theories often sit
at such features of bootstrap bounds. The three-dimensional Ising model, for instance, sits at a kink of the upper bound
on the dimension of its energy operator and at a sharp minimum of the lower bound on its central charge, both viewed as
functions of the dimension of its spin field~\cite{El-Showk:2012cjh,Poland:2018epd}. El-Showk et al.\ therefore
conjectured that it minimizes the central charge among all unitary solutions of crossing
symmetry~\cite{El-Showk:2014dwa}.

\begin{figure}[t]
\centering
\includegraphics[width=\textwidth]{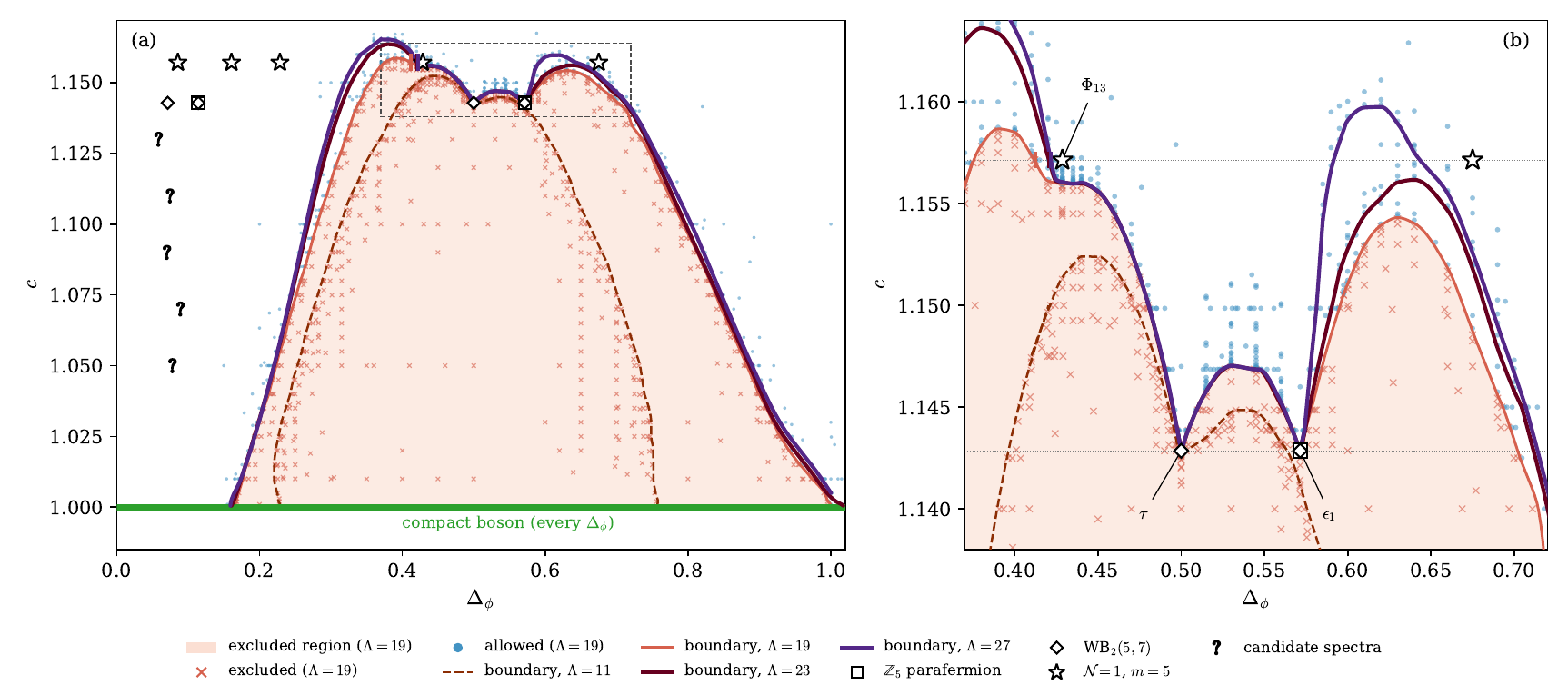}
\caption{(a) Excluded region (shaded) at $\Lambda=19$, with $\Delta_\phi$ on the horizontal axis. Its upper boundary is
$c_{\min}(\Delta_\phi)$. Crosses and dots: excluded and allowed points at $\Lambda=19$. They include points excluded at a lower $\Lambda$
and points allowed at a higher $\Lambda$, since the excluded region can only grow with $\Lambda$. Curves: boundary of the
allowed region from bisection at $\Lambda=11$ (dashed), 19 (solid), 23 and 27 (thick).
Green line: $c=1$, where the compact boson has, as its radius varies, a scalar primary of every dimension
$\Delta_\phi\le2$ that satisfies~\eqref{eq:assumption}. Squares, diamonds
and stars: scalar primaries that satisfy~\eqref{eq:assumption} by their fusion rules in the known theories with
$1<c<6/5$, namely the $\mathbb{Z}_5$ parafermion and $\mathrm{WB}_2(5,7)$ at $c=8/7$ and the $\mathcal{N}=1$ minimal
model with $m=5$ at $c=81/70$. Question marks: lightest scalars of the candidate spectra of Benjamin, Fitzpatrick, Li
and Thaler at $c=1.05$--$1.13$~\cite{Benjamin:2026lbj}. (b) Enlargement of the dashed box in (a), with labels for the energy operator $\epsilon_1$ of the $\mathbb{Z}_5$
parafermion, the twist field $\tau$ of $\mathrm{WB}_2(5,7)$, and the field $\Phi_{13}$ of the $\mathcal{N}=1$ minimal
model.}
\label{fig:map}
\end{figure}

We first look just above $c=1$, where the natural comparison is the compact boson. At $c=1$ the dimension of the operator $\cos(X/R)$ of the boson takes every positive value as the radius $R$
varies~\cite{Ginsparg:1987eb,Dijkgraaf:1987vp}. Since this operator satisfies our assumption for $\Delta_\phi\le2$
(Section~\ref{sec:setup}), no $\Delta_\phi\le2$ can be excluded at $c=1$. Just above $c=1$, in
contrast, we find that crossing symmetry excludes the whole interval $0.17\lesssim\Delta_\phi\lesssim1$
(Figure~\ref{fig:map}). The excluded interval persists as $c$ decreases towards 1. The points that we computed in this
interval at $c=1.00001$ are still excluded (Section~\ref{sec:window}).

Combining the excluded interval with results from the modular bootstrap, we obtain an upper bound on the gap. Modular invariance implies that
the lightest primary of a compact unitary CFT just above $c=1$ is a scalar with dimension below about $1/2$~\cite{Collier:2016cls}.
Since the excluded interval extends well above $1/2$, this scalar must lie below it. We therefore find $\Delta_{\rm gap}<0.17$ at $c=1.0005$ and at $c=1.001$, the upper end of Hellerman's range. This bound is a third of the largest gap possible at
$c=1$ and lower than the modular bound with integer degeneracies, whose minimum is about $0.2$~\cite{Thaler:2026talk}.\footnote{After the first version of this paper appeared,
Benjamin, Langhoff, Li and Thaler showed that integer degeneracies, imposed on the primaries of spin one, lower the
modular bound to $\Delta_{\rm gap}<0.041$ at $c=1.003$~\cite{Benjamin:2026bllt}. Our bound rests on different
assumptions. It uses the crossing equation of a four-point function together with the modular bound without integer
degeneracies.} Our bound weakens as $c$ increases, but at each central charge we computed up to $c=1.10$ (Table~\ref{tab:window}) it
remains below $0.3$, so that a gap in the range $0.3<\Delta_{\rm gap}<0.5$ is impossible there. In this range of the gap,
Benjamin, Fitzpatrick, Li and Thaler found no candidates for $c<1.06$, and they suggested that a stronger bound on the
gap might explain why. The candidates they did find have scalars lighter than our bound. We tested three of them directly, taking $\phi$ to be the lightest scalar of a candidate and allowing in $\phi\times\phi$
only the operators of that candidate. All three are consistent with the crossing equation, one of them only if some of
its dimensions, which are known only numerically, lie outside their quoted uncertainties (Section~\ref{sec:window}).

Further from $c=1$, scalars of the two known theories at $c=8/7$ lie on the boundary of the excluded region. As $c$
increases, the excluded interval shrinks and disappears at $c\approx1.16$. At fixed $\Delta_\phi$ the excluded region therefore ends at a central charge $c_{\min}(\Delta_\phi)$, above which the
point $(c,\Delta_\phi)$ is allowed. As a function of $\Delta_\phi$, $c_{\min}$ is the upper boundary of the shaded region in
Figure~\ref{fig:map}. Below this boundary, every point that we computed at $c\ge1.0005$ is excluded at $\Lambda=19$. Closer
to $c=1$, we did not determine the edges of the excluded interval at this $\Lambda$ (Section~\ref{sec:window}). Hence
$c_{\min}(\Delta_\phi)$ is a lower bound on the central charge of any unitary CFT with $c\ge1.0005$ that has a primary $\phi$
of dimension $\Delta_\phi$ satisfying our assumption. Near the top of the
shaded region, $c_{\min}(\Delta_\phi)$ has two sharp local minima, close to $\Delta_\phi=1/2$ and $4/7$. These two values are exactly the dimensions of scalars in the two
theories at $c=8/7$. The $\mathbb{Z}_5$ parafermion has an energy operator with $\Delta=4/7$, and its
charge-conjugation orbifold, the $W(2,4)$ minimal model $\mathrm{WB}_2(5,7)$, has a twist field with $\Delta=1/2$. As we increase $\Lambda$, $c_{\min}$ at these two values of $\Delta_\phi$ approaches $8/7$, and at the highest order we use, it lies within $2\times10^{-7}$ of $8/7$. In this sense, the two theories at $c=8/7$
minimize the central charge at fixed $\Delta_\phi$, as the three-dimensional Ising model is conjectured to do. We obtain these minima using only Virasoro symmetry, crossing symmetry, unitarity and our assumption on $\phi$. Crossing symmetry also gives an upper bound on the dimension of the lightest
scalar in $\phi\times\phi$ heavier than $\phi$. Both theories saturate this bound as well. At the bound, the crossing
equation truncated to derivatives of order at most $\Lambda$ has a unique solution, and the dimensions in this solution
agree with those of Virasoro primaries in $\phi\times\phi$ of the two theories (Section~\ref{sec:wedges}).

The other scalars of the three known theories that satisfy our assumption all lie inside the allowed region
(Figure~\ref{fig:map}). Among them, the field $\Phi_{13}$ of the $\mathcal{N}=1$ minimal model at $c=81/70$ lies just
above the boundary, next to a short plateau of $c_{\min}$ (Section~\ref{sec:n1}).

Just above $c=1$, the region of light scalars, $\Delta_\phi\lesssim0.17$, remains allowed, but no theory is known
there. Whether it contains one is what remains of Hellerman's question.

\section{Setup}
\label{sec:setup}

We consider the four-point function of a
real scalar Virasoro primary $\phi$ with conformal weights $h_\phi=\bar h_\phi=\Delta_\phi/2$ in a unitary CFT with
$c>1$,
\begin{equation}
\langle\phi(0)\phi(z,\bar z)\phi(1)\phi(\infty)\rangle = \sum_{\mathcal{O}} \lambda_{\phi\phi\mathcal{O}}^2\,
\mathcal{F}_{h_\mathcal{O}}(z)\,\mathcal{F}_{\bar h_\mathcal{O}}(\bar z).
\label{eq:blocks}
\end{equation}
Here the sum runs over the Virasoro primaries $\mathcal{O}\in\phi\times\phi$, with conformal weights
$(h_\mathcal{O},\bar h_\mathcal{O})$, scaling dimension $\Delta_\mathcal{O}=h_\mathcal{O}+\bar h_\mathcal{O}$ and spin
$\ell_\mathcal{O}=|h_\mathcal{O}-\bar h_\mathcal{O}|$, and $\mathcal{F}_h$ are the Virasoro blocks~\cite{Zamolodchikov:1984eqp,Zamolodchikov:1987avt} at central charge
$c$. We take all CFTs to be bosonic, so that all spins are integers. Apart from this, our only assumption about the spectrum is
\begin{equation}
\Delta_\mathcal{O} \ge \max(\Delta_\phi,\ \ell_\mathcal{O}) \qquad \text{for every primary } \mathcal{O}\neq 1
\text{ in } \phi\times\phi ,
\label{eq:assumption}
\end{equation}
that is, no primary lighter than $\phi$ appears in the OPE $\phi\times\phi$ (the condition $\Delta\ge\ell$ is the
unitarity bound). In a
compact CFT the lightest primary satisfies~\eqref{eq:assumption} automatically, and heavier primaries can satisfy it
as well. In the compact boson, for instance, $\cos(X/R)\times\cos(X/R)$ contains, besides the identity, only
$\cos(2X/R)$, with $\Delta=4\Delta_\phi$, and other Virasoro primaries with $\Delta\ge2$, the lightest of which is
$\partial X\bar\partial X$. The operator $\cos(X/R)$ therefore satisfies~\eqref{eq:assumption} whenever
$\Delta_\phi\le2$.

Exchanging the operators at $0$ and $1$ maps $z$ to $1-z$, so that crossing symmetry requires
\begin{equation}
\sum_{\mathcal{O}} \lambda_{\phi\phi\mathcal{O}}^2\, F_{h_\mathcal{O},\bar h_\mathcal{O}}(z,\bar z)=0,\qquad
F_{h,\bar h}(z,\bar z)=\mathcal{F}_h(z)\mathcal{F}_{\bar h}(\bar z)-\mathcal{F}_h(1-z)\mathcal{F}_{\bar h}(1-\bar z),
\label{eq:crossing}
\end{equation}
where the identity contributes $F_{0,0}$ with unit coefficient for a unit-normalized
$\phi$.\footnote{Liouville theory is unitary for every $c>1$~\cite{Ribault:2014hia} and has scalar primaries of every dimension above
$(c-1)/12$. It does not contradict the exclusions below, because it gives no solution of~\eqref{eq:crossing} that
satisfies~\eqref{eq:assumption}. Its primaries are normalizable only as $\delta$-functions, so that their OPE contains no identity. The continuum in this
OPE starts at
$\Delta=(c-1)/12$~\cite{Dorn:1994xn,Zamolodchikov:1995aa,Nakayama:2004vk}, below the dimension of the primaries
themselves.} In a unitary CFT the coefficients $\lambda_{\phi\phi\mathcal{O}}^2$ are non-negative, so that
\eqref{eq:crossing} says that a sum of the functions $F_{h,\bar h}$ with non-negative coefficients cancels the
contribution of the identity. To show that this is impossible at a point $(c,\Delta_\phi)$, we look for a linear
functional $\alpha$ on the derivatives $\partial_z^m\partial_{\bar z}^n$ of~\eqref{eq:crossing} at $z=\bar z=1/2$ with
$m+n\le\Lambda$ such that
\begin{equation}
\alpha[F_{0,0}]>0,\qquad \alpha[F_{h,\bar h}+F_{\bar h,h}]\ge0 \quad\text{for every }(h,\bar h)\text{ allowed
by~\eqref{eq:assumption}}.
\label{eq:functional}
\end{equation}
Here we have combined $(h,\bar h)$ with $(\bar h,h)$, which appear in~\eqref{eq:crossing} with equal
coefficients.\footnote{We do not assume parity. In a parity-invariant CFT, $(h,\bar h)$ and $(\bar h,h)$ appear with equal
coefficients because parity exchanges $h$ and $\bar h$. In any other CFT, the four-point
function averaged over $z\leftrightarrow\bar z$ satisfies~\eqref{eq:crossing} as well, with non-negative coefficients that
are equal for $(h,\bar h)$ and $(\bar h,h)$. Likewise, the modular bound that we use in Section~\ref{sec:window} holds without parity, because the partition function averaged over $h\leftrightarrow\bar h$ is modular invariant as well and
has the same gap~\cite{Collier:2016cls}.} If such
a functional exists, applying it to~\eqref{eq:crossing} gives a contradiction, because the left-hand side becomes positive. No unitary CFT
then has a primary $\phi$ with these values of $c$ and $\Delta_\phi$ that satisfies~\eqref{eq:assumption}. A functional
that excludes a point at order $\Lambda$ also excludes it at every higher order, so that the excluded region can only
grow with $\Lambda$. We
search for these functionals with the semidefinite program solver SDPB~\cite{Simmons-Duffin:2015qma,Landry:2019qug}, as
described in Appendix~\ref{app:numerics}. We first map the $(c,\Delta_\phi)$ plane at $\Lambda=11$ and $19$, and we then determine the parts of the boundary of the excluded region discussed below by bisection up to
$\Lambda=31$.

\section{Just above $c=1$}
\label{sec:window}

Close to $c=1$, the excluded region at fixed $c$ contains an interval $D_\Lambda(c)<\Delta_\phi<U_\Lambda(c)$. We
determine its endpoints by bisection (Table~\ref{tab:window}), and every point that we computed between them is
excluded. The excluded interval at $\Lambda=11$ and $19$ hardly changes as $c$ decreases towards 1. At $\Lambda=11$ its
endpoints at $c=1.0001$ agree with those at $c=1.0005$ within the accuracy of the bisection. At $c=1.00001$, the central
charge closest to 1 at which we computed, the points listed in Table~\ref{tab:window} are excluded at both $\Lambda=11$
and $19$. The upper endpoint grows with
$\Lambda$. At $c=1.001$, $U_{19}=0.999$, whereas $U_{23}=1.015$, so that the
excluded interval extends beyond $\Delta_\phi=1$.

\begin{table}[h]
\centering
\small
\begin{tabular}{c|cc|cc|cc|cc}
\toprule
& \multicolumn{2}{c|}{$\Lambda=11$} & \multicolumn{2}{c|}{$\Lambda=19$} & \multicolumn{2}{c|}{$\Lambda=23$} & \multicolumn{2}{c}{$\Lambda=27$} \\
$c$ & lower & upper & lower & upper & lower & upper & lower & upper \\
\midrule
1.0005 & 0.229 & 0.757 & 0.166 & 0.999 & 0.161 & 1.017 & 0.159 & $>1.055$ \\
1.001 & 0.229 & 0.757 & 0.166 & 0.999 & 0.162 & 1.015 & 0.159 & 1.059 \\
1.01 & 0.221 & 0.752 & 0.177 & 0.974 & 0.174 & 0.980 & 0.174 & 0.987 \\
1.05 & 0.253 & 0.733 & 0.223 & 0.883 & 0.221 & 0.886 & 0.220 & 0.890 \\
1.10 & 0.314 & 0.661 & 0.281 & 0.793 & 0.268 & 0.797 & 0.265 & 0.805 \\
1.14 & 0.389 & 0.576 & 0.334 & 0.714 & 0.312 & 0.720 & 0.305 & 0.723 \\
\midrule
1.0001 & 0.230 & 0.758 & \multicolumn{2}{c|}{excluded: 0.25, 0.5} & & & & \\
1.00001 & \multicolumn{2}{c|}{excluded: 0.3, 0.45, 0.7} & \multicolumn{2}{c|}{excluded: 0.45, 0.9} & & & & \\
\bottomrule
\end{tabular}
\caption{Endpoints of the excluded interval, $D_\Lambda(c)<\Delta_\phi<U_\Lambda(c)$, determined by bisection to
$\pm0.002$ ($\pm0.004$ for the upper endpoint at $c=1.14$, $\Lambda=19$). At $c=1.0005$ and $\Lambda=23$, the points $\Delta_\phi=1.019$, $1.022$, $1.028$ and $1.034$ above the interval
are allowed, whereas the points that we computed between $1.040$ and $1.056$ are excluded. The entry $>1.055$ means
that we did not determine this endpoint. At $\Lambda=27$ the points $\Delta_\phi=1.022$, $1.028$, $1.034$ and $1.037$
are excluded. Since the excluded region can only grow with $\Lambda$, so are all points excluded at $\Lambda=23$, among
them those between $1.040$ and $1.056$. Between the lower endpoint and the modular bound $\Delta_{\rm mod}(c)$ defined in the text, every $\Delta_\phi$ we computed is excluded, on grids of spacing at most
$0.07$. For the bound \eqref{eq:gap} we use the lowest points excluded at $\Lambda=19$, $\Delta_\phi=0.1668$ at $c=1.0005$
and $0.1661$ at $c=1.001$. At $c=1.0001$ we determined the endpoints only at $\Lambda=11$, and at $\Lambda=19$, as at
$c=1.00001$, we computed only the points listed in the last two rows.}
\label{tab:window}
\end{table}

We now combine the excluded interval with results from the modular bootstrap. Modular invariance implies an upper bound
$\Delta_{\rm mod}(c)$ on the dimension of the lightest primary. This bound rises from $1/2$ at $c=1$ but stays below $0.525$ up to
$c=1.14$~\cite{Collier:2016cls}. The lightest primary therefore has $\Delta<1$, and since spins in a bosonic CFT are
integers that cannot exceed the dimension, it is a scalar. As the lightest primary of the theory, it
satisfies~\eqref{eq:assumption} (for a complex primary, its real part does), so that its dimension cannot lie inside
the excluded interval. Nor can it lie above the excluded interval, because $\Delta_{\rm mod}(c)<U_{19}(c)$. Hence
\begin{equation}
\Delta_{\rm gap} < D_\Lambda(c)
\label{eq:gap}
\end{equation}
in any compact, unitary, bosonic CFT, at each value of $c$ from $1.0005$ to $1.14$ in Table~\ref{tab:window}. For this conclusion we need the whole range between $D_\Lambda(c)$ and $\Delta_{\rm mod}(c)$ to be excluded, which we
checked on grids of points at each of these values of $c$ (Table~\ref{tab:window}). At $c=1.0005$ and $1.001$, the lowest points whose exclusion at $\Lambda=19$ we verified independently (Appendix~\ref{app:numerics}) are
$\Delta_\phi=0.1668$ and $0.1661$, so that $\Delta_{\rm gap}<0.167$ at both central charges. These bounds improve at higher derivative order, since the lower endpoint decreases as $\Lambda$ increases (Table~\ref{tab:window}).

The bound~\eqref{eq:gap} is of no use for the candidate spectra of Benjamin, Fitzpatrick, Li and Thaler at
$c=1.05$--$1.13$~\cite{Benjamin:2026lbj}, because their lightest scalars, with dimensions between $0.059$ and $0.089$, lie well below the excluded interval
(Figure~\ref{fig:map}a). Instead, we tested the spectra themselves. Taking $\phi$ to be the lightest scalar of a candidate, we allowed in
$\phi\times\phi$ only the operators of that candidate, each in a small interval around its quoted dimension
(Appendix~\ref{app:numerics}). We did this for the candidates at $c=1.05$, $1.09$ and $1.13$. The other two, at $c=1.07$
and $1.11$, have several low-lying dimensions that are uncertain by order one, so that this procedure would hardly constrain them. To
check that the procedure can exclude a wrong spectrum, we applied it to the tensor product of the tricritical Ising model and the Ising model,
with $\phi$ the energy operator of the tricritical Ising model. The actual spectrum of $\phi\times\phi$ in this theory is allowed at $\Lambda=11$ and $19$,
whereas the same spectrum with one scalar moved by $0.1$ is excluded. All three candidates are consistent with crossing symmetry, but the one at $c=1.05$ is consistent at $\Lambda=19$ only if some of its
dimensions lie outside their quoted uncertainties.

\section{Two theories at $c=8/7$}
\label{sec:wedges}

\begin{figure}[t]
\centering
\includegraphics[width=0.44\textwidth]{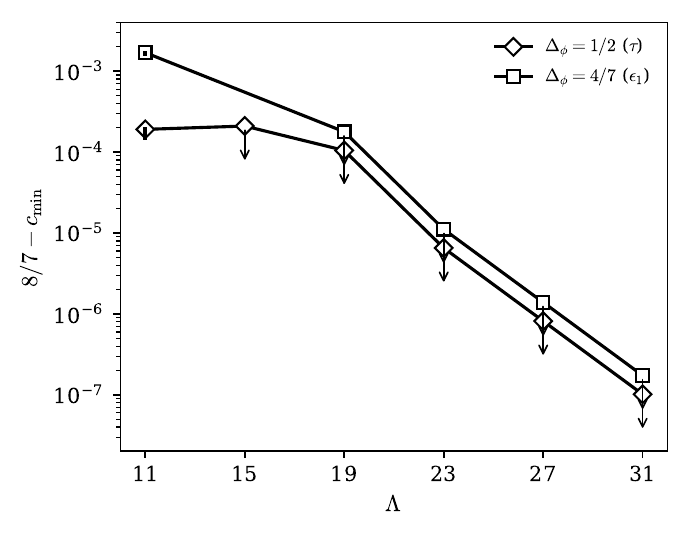}
\caption{Distance of $c_{\min}(1/2)$ and $c_{\min}(4/7)$ from $8/7$ as a function of the derivative order $\Lambda$.
We determine $c_{\min}$ by bisection in $c$. Since $c=8/7$ itself is allowed, a bisection in which every value of $c$
that we tested below $8/7$ is excluded tells us only that $c_{\min}$ lies between the largest excluded value and $8/7$.
The corresponding point is then an upper limit (arrow).}
\label{fig:tips}
\end{figure}

In this section we show that the two minima of $c_{\min}$ near $\Delta_\phi=1/2$ and $4/7$ (Figure~\ref{fig:map}b)
correspond to the two known theories at $c=8/7$. Evaluated at exactly these two values, $c_{\min}$ approaches $c=8/7$ from below as
$\Lambda$ increases, and at $\Lambda=31$ it lies within $2\times10^{-7}$ of $8/7$ (Figure~\ref{fig:tips}). Since the two theories
exist, the point $c=8/7$ can never be excluded at these two values of $\Delta_\phi$.

The minimum at $\Delta_\phi=4/7$ corresponds to the $\mathbb{Z}_5$ parafermion. In the description of the $\mathbb{Z}_k$ parafermion as the coset $SU(2)_k/U(1)$, the energy operators $\epsilon_j$
carry $SU(2)$ spin $j$ and no $U(1)$ charge, and they have $h=\bar h=j(j+1)/(k+2)$~\cite{Fateev:1985mm}. For $k=5$,
the operator $\epsilon_1$ therefore has $\Delta=4/7$, and $\epsilon_2$ has $\Delta=12/7$. Although $\epsilon_1$ is not the lightest primary of
the theory (the spin field $\sigma_1$ has $\Delta=4/35$), it satisfies~\eqref{eq:assumption}. Since $\epsilon_1$ carries no
$\mathbb{Z}_5$ charge, the fusion rules allow in $\epsilon_1\times\epsilon_1$ only neutral primaries from the
representations of the parafermion algebra built on the identity, on $\epsilon_1$ and on $\epsilon_2$. All of them except the identity have $\Delta\ge4/7$. Moreover, the coefficient of $\epsilon_1$ itself vanishes, since it contains the $SU(2)$ $3j$ symbol $\left(\begin{smallmatrix}1&1&1\\0&0&0\end{smallmatrix}\right)=0$ as a factor, so that the OPE
$\epsilon_1\times\epsilon_1=1+\epsilon_2+\dots$ starts with $\epsilon_2$. Because assumption~\eqref{eq:assumption} allows $\epsilon_1$ in its own OPE, the minimum at $\Delta_\phi=4/7$ appears
without this selection rule as input.

The minimum at $\Delta_\phi=1/2$ corresponds to the charge-conjugation orbifold of the $\mathbb{Z}_5$ parafermion, the
minimal model $\mathrm{WB}_2(5,7)$ of the $W(2,4)$
algebra~\cite{Kausch:1990bn,Blumenhagen:1990jv,Jiang:2019ghv,Gepner:2013qca}. Its twisted sector contains a field
$\tau$ with $h=\bar h=1/4$~\cite{Jiang:2019ghv}. Like $\epsilon_1$, the field $\tau$ is not the lightest primary of its
theory, since the twisted sector also contains fields with $\Delta=1/14$ and $3/14$. To see that $\tau$ nevertheless
satisfies~\eqref{eq:assumption}, we computed the fusion rules of the orbifold from its modular $S$-matrix with the Verlinde formula~\cite{Verlinde:1988sn}. In terms of representations of the $W(2,4)$ algebra, they give $\tau\times\tau=1+[\psi_1\bar\psi_1]+[\psi_2\bar\psi_2]$.
Here $1$ denotes the vacuum representation, which contains the currents of the algebra, and $[\psi_j\bar\psi_j]$ the
representation built on the charge-conjugation-invariant combination of $\psi_j\bar\psi_j$ and its conjugate, where
$\psi_j$ are the parafermion currents with $h=j(5-j)/5$. The lightest Virasoro primary in $\tau\times\tau$ other than the
identity therefore has $\Delta=8/5$, well above $\Delta_\tau=1/2$.

Besides lying at the minima of $c_{\min}$, both theories saturate a second bound. At a point $(c,\Delta_\phi)$, we require every scalar in $\phi\times\phi$ other than the identity to have $\Delta\ge g$,
and we raise $g$ until the point is excluded. Since $g>\Delta_\phi$, this requirement also removes $\phi$ itself from
$\phi\times\phi$. For $\tau$, the removal agrees with the fusion rules of the orbifold, which exclude $\tau$ from
$\tau\times\tau$. For $\epsilon_1$, we nevertheless allow $\epsilon_1$ itself in its own OPE, because the fusion rules of the
parafermion algebra permit it and we do not impose the selection rule that sets its coefficient to zero. The largest $g$
for which the point is still allowed is an upper bound $\Delta_*$ on the dimension of the lightest scalar in
$\phi\times\phi$ heavier than $\phi$. At $c=8/7$, this bound agrees with the exact dimension of that scalar in the two theories, $8/5$ for $\tau$ at
$\Delta_\phi=1/2$ and $12/7$ for $\epsilon_1$ at $\Delta_\phi=4/7$, to within $10^{-3}$ (Table~\ref{tab:spectra}).
At such a bound, the truncated crossing equation has a unique solution, and the zeros of the extremal functional, that
is, of the functional~\eqref{eq:functional} at the bound, are the dimensions in this
solution~\cite{El-Showk:2012vjm}. Here the zeros lie within $3\times10^{-3}$ of the dimensions of Virasoro primaries
of the two theories (Table~\ref{tab:spectra}). The converse does not hold, because the functional has no zero at the
spin-4 current of the $W(2,4)$ algebra, which appears in $\tau\times\tau$.

\begin{table}[t]
\centering
\footnotesize
\setlength{\tabcolsep}{4pt}
\begin{tabular}{lll}
\toprule
& $\tau$ in $\mathrm{WB}_2(5,7)$, $\Delta_\phi=1/2$ & $\epsilon_1$ in $\mathbb{Z}_5$, $\Delta_\phi=4/7$ \\
\midrule
$\Delta_*$ at $c=8/7$ ($\Lambda=11$ and 19) & $1.600<\Delta_*<1.601$ & $1.714<\Delta_*<1.715$ \\
exact & $8/5$ ($\psi_1\bar\psi_1$) & $12/7=1.7143$ ($\epsilon_2$) \\
\midrule
$\Lambda=19$ zeros, $\ell=0$ & 2.4012 [12/5], 6.4018 [32/5], 8.0019 [8] & 2.5723 [18/7], 6.5739 [46/7], 8.0023 [8] \\
$\ell=2$ & 4.4000 [22/5] & 4.5714 [32/7], 8.5732 [60/7] \\
$\ell=4$ & 5.6000 [28/5] & 5.7143 [40/7], 6.5714 [46/7] \\
$\ell=6$ & 7.6001 [38/5] & 8.5716 [60/7] \\
$\ell=8$ & 8.0000 [8], 9.5999 [48/5] & 8.0000 [8], 9.7145 [68/7] \\
\bottomrule
\end{tabular}
\caption{The theories at the two minima of $c_{\min}$. Top: the upper bound $\Delta_*$ on the scaling dimension of the lightest scalar primary in $\phi\times\phi$ heavier
than $\phi$, and the exact dimension of this scalar. Bottom: zeros of the extremal functional with
$\Delta<10$ at the bound at $\Lambda=19$ (spins $\ell\le8$), with the dimensions of the corresponding Virasoro primaries
of the exact theory in brackets. For $\epsilon_1$ we computed the bound with $\epsilon_1$ itself included in the spectrum of $\phi\times\phi$, as explained
in the text. At $\Lambda=11$ the bound is the same when $\epsilon_1$ is left out, so that the saturation does not rely on
the selection rule.}
\label{tab:spectra}
\end{table}

\section{The $\mathcal{N}=1$ minimal model}
\label{sec:n1}

The third known theory above $c=1$ is the $\mathcal{N}=1$ minimal model with $m=5$, at $c=81/70$. Its Neveu--Schwarz
superconformal primaries $\Phi_{r,s}$ have $h=\bar h=\bigl((7r-5s)^2-4\bigr)/280$, so that $\Phi_{13}$ has
$\Delta=3/7$. To decide whether $\Phi_{13}$ satisfies~\eqref{eq:assumption}, we need the Virasoro primaries in
$\Phi_{13}\times\Phi_{13}$, since our crossing equation is written in terms of Virasoro blocks. The superconformal family
of each $\Phi_{r,s}$ contains several Virasoro primaries: $\Phi_{r,s}$ itself, the top component
$G_{-1/2}\bar G_{-1/2}\Phi_{r,s}$, whose dimension is larger by one, and further Virasoro primaries of higher dimension.
The OPE $\Phi_{13}\times\Phi_{13}$ contains Virasoro primaries from the superconformal families of the identity, of
$\Phi_{13}$ and of $\Phi_{15}$. From the superconformal family of $\Phi_{13}$, however, only the top component and the Virasoro
primaries above it appear, and $\Phi_{13}$ itself does not~\cite{Alvarez-Gaume:1991bj}. The lightest Virasoro primary
in $\Phi_{13}\times\Phi_{13}$ other than the identity is therefore the top component of $\Phi_{13}$, with
$\Delta=10/7$, followed by $\Phi_{15}$ with $\Delta=16/7$, so that $\Phi_{13}$ satisfies~\eqref{eq:assumption}.

Like the two theories at $c=8/7$, this model saturates the upper bound $\Delta_*$ on the dimension of the lightest scalar
in $\phi\times\phi$ heavier than $\phi$. Since $\Phi_{13}$ does not appear in its own OPE, we compute this bound without
it. At $(c,\Delta_\phi)=(81/70,3/7)$, the bound lies within $5\times10^{-4}$ of $10/7$, the dimension of the top
component.
Unlike $\tau$ and $\epsilon_1$, however, $\Phi_{13}$ does not lie on the boundary $c_{\min}$ but just above it. Just
below this point, $c_{\min}$ has a short plateau that does not move between $\Lambda=19$ and $27$ (Figure~\ref{fig:map}b).

On the plateau, the solution of the crossing equation, which we read off from the zeros of the extremal functional, is
nevertheless close to the OPE $\Phi_{13}\times\Phi_{13}$ of the $\mathcal{N}=1$ minimal model. In this OPE, apart from the conserved currents in the superconformal family of the identity, the Virasoro primaries
of the two lowest twists $\Delta-\ell$ belong to the superconformal families of $\Phi_{13}$, at twist $10/7$, and of
$\Phi_{15}$, at twist $16/7$. On the plateau, the zeros
lie at the same two twists, each shifted by about $10^{-3}$, and at the same spins, except at $\ell=6$
(Table~\ref{tab:n1}). The solution on the plateau thus has the structure of the OPE $\Phi_{13}\times\Phi_{13}$, although no
$\mathcal{N}=1$ minimal model exists at its central charge.

\begin{table}[t]
\centering
\small
\begin{tabular}{c|cc|cc}
\toprule
& \multicolumn{2}{c|}{twist $\approx10/7$} & \multicolumn{2}{c}{twist $\approx16/7$} \\
$\ell$ & plateau, $\Lambda=27$ & $\mathcal{N}=1$ & plateau, $\Lambda=27$ & $\mathcal{N}=1$ \\
\midrule
0 & 1.42991 & $10/7=1.42857$ & 2.28435 & $16/7=2.28571$ \\
2 & 3.42992 & $24/7=3.42857$ & 4.28434 & $30/7=4.28571$ \\
4 & 5.42992 & $38/7=5.42857$ & -- & -- \\
6 & -- & $52/7=7.42857$ & 8.28470 & $58/7=8.28571$ \\
8 & 9.42990 & $66/7=9.42857$ & 10.28432 & $72/7=10.28571$ \\
\bottomrule
\end{tabular}
\caption{Zeros of the extremal functional with spin $\ell\le8$ on the plateau of $c_{\min}$ at $(c,\Delta_\phi)=(1.15604,3/7)$, compared
with the Virasoro primaries of twist $10/7$ and $16/7$ in $\Phi_{13}\times\Phi_{13}$ of the $\mathcal{N}=1$ minimal
model ($c=81/70$), which belong to the superconformal families of $\Phi_{13}$ and $\Phi_{15}$, respectively. A dash
means that there is no such operator. In particular, neither spectrum has an operator of twist $16/7$ at $\ell=4$. At $\ell=6$ the functional has no zero near $\Delta=52/7$, the dimension of the operator of twist $10/7$ in the
$\mathcal{N}=1$ minimal model. Its squared OPE coefficient, which we computed from the $\mathcal{N}=1$ superconformal Ward identities, is tiny,
$2.7\times10^{-10}$. That of the operator of twist $16/7$ at the same spin, $8.2\times10^{-10}$, is comparable, however,
and this operator does have a zero. Further zeros with $\Delta<10$ lie at $\Delta=4$ ($\ell=4$) and $8$ ($\ell=8$), where the $\mathcal{N}=1$ minimal
model has conserved currents, at $\Delta=6.283$ and $8.000$ for $\ell=0$ (exact: $44/7=6.2857$ and 8), and at $\Delta=7.426$ for $\ell=2$ (exact: $52/7=7.4286$). The zeros at $\Delta=5.405$ ($\ell=0$), $5.962$ and
$9.948$ ($\ell=2$), $4.532$ ($\ell=4$) and $8.413$ ($\ell=8$) have no counterpart within $10^{-2}$.}
\label{tab:n1}
\end{table}

\section{Discussion}
\label{sec:disc}

We have found that just above $c=1$, crossing symmetry excludes every $\Delta_\phi$ between about $0.17$ and $1$,
although at $c=1$ the compact boson has, for a suitable radius, a scalar of each of these dimensions that satisfies our
assumption. If the excluded interval reaches $c=1$ as $\Lambda\to\infty$, the allowed region is therefore
discontinuous at $c=1$. Together with results from the modular bootstrap, the excluded interval implies that a compact unitary CFT
at $c=1.0005$ or $1.001$ must have a scalar lighter than $0.17$. The candidate spectra of Benjamin, Fitzpatrick, Li and
Thaler at $c=1.05$--$1.13$, outside Hellerman's range, have their lightest scalars below the excluded interval, as
\eqref{eq:gap} requires. To test them further, one would need their OPE coefficients, which the partition function does
not determine.

We expect that by also using the four-point functions that involve the lightest scalar in $\phi\times\phi$ heavier
than $\phi$, one can isolate the two theories at $c=8/7$ in small islands in the space of $c$, $\Delta_\phi$ and
the dimension of that second scalar, as one can for the three-dimensional Ising model~\cite{Kos:2014bka}. With these
four-point functions we also hope to constrain the region of light scalars further and to understand the plateau of
$c_{\min}$ just below the $\mathcal{N}=1$ minimal model.

To come back to Hellerman's question, with which we began, the author's answer is still yes. At the upper end of his range,
$c=1.001$, crossing symmetry and modular invariance now forbid every compact unitary CFT whose lightest primary is
heavier than $0.17$. If everything not forbidden is compulsory, a theory with a lighter scalar must exist there.

\section*{Acknowledgments}

The author thanks Simeon Hellerman for the question that started this work. The author shared this question with
Alessandro Vichi at the workshop ``Progress of Theoretical Bootstrap'' at YITP, Kyoto, in 2025. Stefanos Kousvos,
Alessandro Piazza and Vichi have studied it independently~\cite{Kousvos:2026vb}, and the author thanks them for
coordinating the submission of the two papers to arXiv. The code for the Virasoro conformal bootstrap in this work was written by Claude Opus 5.5
(Anthropic), which also ran the numerical computations and assisted in drafting the manuscript. This work is supported
in part by JSPS KAKENHI Grant Numbers 21K03581 and 26K00699.

\appendix
\section{Numerical details}
\label{app:numerics}

SDPB requires the positivity conditions~\eqref{eq:functional} to be written as positivity of polynomials in the conformal weights. We
obtain this form from Zamolodchikov's recursion for the Virasoro blocks in the elliptic nome~\cite{Zamolodchikov:1987avt},
with which each Taylor coefficient of a block at $z=1/2$ has the form $(16q_0)^h N(h)/D(h)$ with $q_0=e^{-\pi}$, where $N(h)$ is a
polynomial and $D(h)=\prod_{r,s}(h-h_{r,s}(c))$ is the product over the poles kept in the recursion. These poles lie at
the zeros $h_{r,s}(c)=Q^2/4-(rb+s/b)^2/4$ of the Kac determinant~\cite{Kac:1978ge}, with $c=1+6Q^2$ and $Q=b+b^{-1}$.
For $1<c<25$ the poles with $r\neq s$ come in complex-conjugate pairs, while those with $r=s\ge2$ lie at $h<0$. The
pole at $h_{1,1}=0$ has vanishing residue because the external operators are identical. Hence $D(h)>0$ for $h\ge0$, so
that positivity of the functional indeed reduces to positivity of polynomials. We truncate the recursion at an order in $q$ between $20$ and $46$,
chosen at each point such that all Taylor coefficients up to order $\Lambda$ agree with those of a table of much
higher order to a relative accuracy of $10^{-20}$ ($10^{-16}$ in the bisections at $\Lambda=11$) for internal conformal weights between $0$ and $40$. We need the
highest orders for $c-1\lesssim10^{-4}$, because as $c\to1^+$ the poles with $r\neq s$ approach the real axis at
$h=(r-s)^2/4$.

To decide whether a point is excluded, we expand the crossing equation in $\partial_z^m\partial_{\bar z}^n$ with $m<n$,
$m+n$ odd and $m+n\le\Lambda$, which gives $55$ components at $\Lambda=19$ and $136$ at $\Lambda=31$. We impose the
positivity conditions~\eqref{eq:functional} for the primaries allowed by~\eqref{eq:assumption}: for even spins
$\ell\le40$ as functions of $\Delta\ge\max(\Delta_\phi,\ell)$, and for $\ell>40$ on 28 lines of fixed twist
$\Delta-\ell\in\{0$, $0.005$, $0.02$, $0.05$, $0.1$, $0.2$, $0.35$, $0.5$, $0.75$, $1$, $1.5$, $2$, $3$, $4$, $6$, $8$, $10$,
$12$, $15$, $18$, $22$, $27$, $33$, $40$, $50$, $70$, $100$, $150\}$, treating $\ell$ as a continuous variable. In
addition, we impose them in the limit $\ell\to\infty$ at fixed twist and in the limit $h,\bar h\to\infty$ at fixed
ratio $\bar h/h$, so that the functional is positive at large spin and dimension as well. We run SDPB with 512 bits of precision and with error thresholds of $10^{-30}$ for the dual and $10^{-60}$ for the primal problem. We count a point as excluded when we obtain a functional that satisfies~\eqref{eq:functional}, and as allowed when we obtain instead a solution of the truncated crossing equation. We determine
the endpoints in Table~\ref{tab:window} by bisection in $\Delta_\phi$ and the values of $c_{\min}$ in Figure~\ref{fig:tips} by bisection in $c$. We read off the spectra in Tables~\ref{tab:spectra} and~\ref{tab:n1} from the zeros of the functional at the
excluded point closest to the bound in question, $\Delta_*$ in Table~\ref{tab:spectra} and $c_{\min}$ in
Table~\ref{tab:n1}.

To test the candidate spectra in Section~\ref{sec:window}, we allow below $\Delta=3.5$ only the operators of the
candidate spectrum. We place each of them in an interval around its quoted dimension, of about the size of the quoted
uncertainty, which we sample with spacing at most $0.005$. Above $\Delta=3.5$ we impose only~\eqref{eq:assumption},
because the spectra are truncated at $\Delta=4$ and are poorly determined near the truncation.  Only for the candidate at $c=1.05$ does the outcome depend on the widths of these intervals. At
$\Lambda=19$ it is excluded, and it remains excluded when we lower the cutoff from $\Delta=3.5$ to $\Delta=3$. It
becomes allowed when we make the intervals $1.5$ times wider.

Because the bound $\Delta_{\rm gap}<0.167$ rests on the exclusion of $\Delta_\phi=0.1668$ at $c=1.0005$, we recomputed
this exclusion with 768 bits of precision and evaluated the resulting functional with blocks from an independent table
of higher order. We found it positive on a grid of even spins up to 3000 and twists up to 1000. Since the poles $h_{r,s}$ come close to the real axis as $c\to1^+$, we also evaluated the functional on finer grids
near the points where one of the two conformal weights equals $(r-s)^2/4$. Taking $h\le\bar h$, so that the twist is $2h$ and the spin is $\ell=\bar h-h$, we used grids of spacing $10^{-3}$ in
twist in two cases. When $h=(r-s)^2/4$, the twist is $(r-s)^2/2$, and we covered these twists up to $12.5$ for spins up
to 3000. When $\bar h=(r-s)^2/4$, the twist is $(r-s)^2/2-2\ell$, and we covered these twists up to 200 for spins up to
400. The functional is positive on all of these grids. We made the same checks at $\Lambda=19$ for the exclusion of
$\Delta_\phi=0.1661$ at $c=1.001$, on which the bound at the upper end of Hellerman's range rests, and for the exclusion of
$\Delta_\phi=0.45$ at $c=1.00001$, in the second case with the grid around the twists
$(r-s)^2/2$ refined to a spacing of $2\times10^{-4}$. In both cases the functional is again positive on all of the grids.

In the first version of this paper, the polynomials that we passed to SDPB for the spins close to $\ell=40$ were
computed with too low a working precision. Their relative error reached $10^{-3}$ at $\ell=40$ near twist $1/2$,
whereas the polynomials on the lines of fixed twist were accurate. Close to $c=1$, where the functionals are positive
only by a small margin, this error changed the outcome at some of the points that we had found to be allowed. We have
therefore recomputed every entry of Table~\ref{tab:window} with polynomials of sufficient precision. Every change
enlarges the excluded region. In particular, the points at $c=1.00001$ that were allowed at $\Lambda=11$, and the point
$\Delta_\phi=0.9$ at $\Lambda=19$, are excluded.

In this version we have also evaluated with an independent table, as above, every functional on which an exclusion in
Table~\ref{tab:window} or the bound~\eqref{eq:gap} rests. Some of the functionals obtained with the positivity
conditions described above are negative at spins above 40, between the lines of fixed twist. For these points we
imposed the conditions also at the spins where the functional is negative, or at all even spins from 42 to 120
together with eight further lines with twists between $200$ and $2300$, and solved again until the functional was positive wherever
we evaluated it. Moreover, the functionals that we obtain satisfy the condition on the identity only up to
the error threshold of SDPB, so that the value of a functional on the identity can be negative at the level of this error. In that case we added to the
functional a small multiple of the vector that the identity contributes to~\eqref{eq:crossing}. This makes the value
on the identity positive and changes the normalized values on all other operators by less than their minimum, so that
the resulting functional excludes the point.

\renewcommand{\bibfont}{\small}
\setlength{\bibsep}{2pt plus 1pt}
\bibliographystyle{unsrtnat}
\bibliography{refs}

\end{document}